**TITLE:**

Comprehensive characterization and validation of a fast-resolving (1000 Hz) plastic scintillator for ultra-high dose rate electron dosimetry

**AUTHORS:**

Lixiang Guo[1*], Banghao Zhou[1], Yi-Chun Tsai[1], Kai Jiang[2], Viktor Iakovenko[2], and Ken Kang-Hsin Wang[1*]

[1]Biomedical Imaging and Radiation Technology Laboratory (BIRTLab), Department of Radiation Oncology, University of Texas Southwestern Medical Center, Dallas, Texas, USA

[2]Department of Radiation Oncology, University of Texas Southwestern Medical Center, Dallas, Texas, USA

**\*CORRESPONDENCE:**

Lixiang Guo

Department of Radiation Oncology, University of Texas Southwestern Medical Center, 5323 Harry Hines Blvd., Dallas, TX 75390, USA

Tel: 469-236-9619

Email: lixiang.guo@utsouthwestern.edu

Ken Kang-Hsin Wang, PhD

Department of Radiation Oncology, University of Texas Southwestern Medical Center, 5323 Harry Hines Blvd., Dallas, TX 75390, USA

Tel: 614-282-0859

Email: kang-hsin.wang@utsouthwestern.edu

**RUNNING TITLE:**

Fast-resolving scintillator for UHDR electron dosimetry

**ABSTRACT**

**Background:** The normal tissue sparing effect of ultra-high dose rate irradiation (≥40 Gy/s, UHDR), as compared to conventional dose rate (CONV), has attracted significant research interest for FLASH radiotherapy (RT). Accurate, dose rate independent, fast-responding dosimeters capable of resolving the spatiotemporal characteristics of UHDR beams are urgently needed to facilitate FLASH research and support its clinical translation. Tissue-equivalent scintillators, with millimeter-level spatial resolution and millisecond-level temporal resolution, possess these required characteristics and show strong potential for use in UHDR dosimetry.

**Purpose:** We investigated the performance of the HYPERSCINT RP-FLASH scintillator system at up to 1000 Hz sampling frequency ($f_s$) for UHDR electron beam dosimetry.

**Methods:** The scintillator was exposed to CONV and UHDR electron irradiation using a LINAC-based FLASH platform. Its spectral characteristics were delineated with a four-component calibration, followed by a signal-to-dose calibration using 18 MeV CONV electron beam. The dose linearity and dosimetric accuracy in response to CONV and UHDR irradiation at 1 and 1000 Hz $f_s$ were quantified against ion chamber and EBT-XD film measurements. The response of the scintillator system was investigated as a function of beam energy (6 and 18 MeV), field size (2x2 to 25x25 cm$^2$), dose per pulse (DPP, 0.8 to 2.3 Gy/pulse), and pulse repetition frequency (PRF, 30 to 180 Hz). Relative signal sensitivity was quantified against accumulated dose to account for the scintillator's radiation degradation. Pulse-resolved dose measurements at 18 MeV UHDR, obtained using the scintillator with 1000 Hz $f_s$ for a train of 10 pulses at 180 Hz PRF, were validated with a PMT-fiber optic scattered radiation detector.

**Results:** The scintillator system at 1 Hz $f_s$ demonstrated high accuracy in dose measurements, remaining within 0.5% of ion chamber measurements over the dose range of 0.1 to 35 Gy under CONV irradiation. For the UHDR irradiation, the scintillator showed <3% dose error compared to film measurements up to 40 Gy at 1000 Hz $f_s$. Its response was found to be minimally dependent on energy, field size, DPP, and PRF. The radiation degradation of the scintillation detector followed a 2nd-order polynomial fit between 0 and 10 kGy, and a linear fit with a slope of -2.6%/kGy in the range of 0 to 2 kGy. The pulse-resolved dose measured by the scintillator was verified to be within 3% accuracy when compared to the measurements obtained using the PMT-fiber optic detector.

**Conclusions:** With routine dose calibration to account for radiation induced degradation, the fast-responding scintillator system can accurately provide millisecond-resolved inter-pulse measurements for electron beams at conventional and ultra-high dose rates, with minimal dependence on beam parameters. This suggests that the HYPERSCINT RP-FLASH scintillator system could serve as a detector of choice for electron FLASH research.

## 1. INTRODUCTION

Ultra-high dose rate ($\geq$40 Gy/s, UHDR) irradiation has been shown to reduce normal tissue toxicity while maintaining similar tumor control compared to conventional dose rate (~0.1 Gy/s, CONV) setting across various animal models and endpoints[1-4]. The differential effect of UHDR between normal tissue and tumor is termed the FLASH effect. The clinical translation of FLASH radiotherapy (RT) has attracted significant interest due to its potential to improve therapeutic ratios, reduce treatment time, and mitigate organ motion-related uncertainties[4-6].

The introduction of UHDR radiation presents new dosimetry challenges[7,8]. Ionization chambers (IC) serve as the clinical reference dosimetry tools for CONV radiation, but most of them exhibit dose rate dependent ion recombination effects at UHDR[9]. Passive detectors, such as radiochromic film and thermoluminescent detector, show dose rate independence and have been applied in UHDR measurements[10,11]. However, these passive detectors require post-irradiation waiting periods, preventing immediate reporting of measurements. In addition to traditional spatial dosimetric measurements like percent depth dose (PDD) and profile, FLASH studies introduce new requirements for resolving temporal beam characteristics. The FLASH effect is observed only above a threshold of average dose rate (ADR)[2] and may depend on other temporal dosimetric factors such as instantaneous dose rate (IDR), pulse interval, and dose per pulse (DPP)[12,13]. Therefore, there is a critical need for a detector that is dose rate independent, fast-responding, and capable of accurately resolving the spatial and temporal characteristics of UHDR beams.

Plastic scintillation detectors (PSDs) have been widely used in CONV because of their favorable dosimetric properties[14,15]. As fast-responding and water-equivalent dosimeters, PSDs have demonstrated energy independence for clinical electron and photon beams, as well as independence on the incident angle in CONV setting[15,16]. PSDs also offer the advantages of high spatial resolution at the millimeter scale and fast temporal resolution on the order of milliseconds[14]. Furthermore, their flexibility in design, such as small size, and magnetic field independence make PSDs suitable for small field[17-21] and MR-linear accelerator (LINAC) dosimetry[22].

A large portion of FLASH studies and clinical trials have involved electron beams[4]. UHDR electron beams are relatively more accessible than other radiation modalities due to methods developed for converting clinic LINACs to UHDR mode[23-26] and various commercial options[27-29]. Beyond plastic scintillators' use in CONV RT, efforts have been made to apply them to address electron FLASH dosimetry needs[30-34]. Liu et al.[31] characterized the Exradin

W2 system (Standard Imaging, Middleton, WI) with a BCF-12 scintillator, demonstrating average dose rate independence and linearity as a function of integrated dose and DPP for DPP ≤ 1.5 Gy and pulse repetition frequency (PRF) ≤ 90 Hz. Oh et al.[32] further characterized the W2 system with a 1 mm diameter x 1 mm length scintillator paired with MAX SD optical detection and signal processing unit, demonstrating within 2% accuracy for a single pulse up to 3.6 Gy, and no discernible dependence on PRF ranging from 18 to 180 Hz. Recently, Baikalov et al.[34] showed that HYPERSCINT RP-FLASH scintillator system exhibited dose linearity independent of wide range of dose rates and DPPs.

To further advance the knowledge in our field, we conducted a comprehensive evaluation of the HYPERSCINT RP-FLASH scintillation dosimetry system, which is specifically designed for UHDR research applications and offers a sampling frequency of up to 1000 Hz. Our study assessed dose linearity and dosimetric accuracy for both CONV and UHDR electron beams, alongside the detector's response to key parameters including beam energy, field size, DPP, PRF and accumulated dose, covering irradiation conditions typically encountered in electron FLASH studies. Pulse-resolved dosimetric measurements of the scintillator system were also validated at 1000 Hz sampling frequency.

## 2. METHODS

### 2.1 Configuration and calibration of HYPERSCINT RP-FLASH scintillator system

The HYPERSCINT RP-FLASH system comprises the HYPERSCINT RP-FLASH optical reader (Fig. 1a1), PRB-0042 RP detector (Fig. 1a2-3), and HYPERDOSE software (Version 0.1.15). The HYPERSCINT RP-FLASH optical reader incorporates a spectrometer with an electric cooled 2D photodetector array for light spectrum measurements. The spectral response of the photodetector was not provided by the vendor and was therefore not applied to the spectral data. While the pixel index of the photodetector corresponds to wavelength, the exact relationship between them was not available. To ensure consistency with the HYPERDOSE software display, we adopted wavelength in arbitrary units (wavelength (a.u.)) as the figure label for the spectra presented in this work. The PRB-0042 RP detector is a single channel plastic scintillator probe with a 1 mm diameter x 3 mm length sensitive volume, coupled to a 20 m long polymethyl methacrylate (PMMA) plastic fiber optic. The reader is connected to a computer through USB 3.0 port. The total dose, temporal dose distribution at selected sampling frequency ($f_s$) and measured light spectrum are saved after each measurement. The system functions at $f_s$ from 1 to 1000 Hz. If $f_s$ < 25 Hz, the system records data continuously until terminated by user. Otherwise, if 25 Hz ≤ $f_s$ ≤ 1000 Hz, the system records a fixed number of sampling windows with a length of $1/f_s$ [34]. The number of sampling windows is 3000 when $f_s$ = 1000 Hz, corresponding to 3s acquisition time.

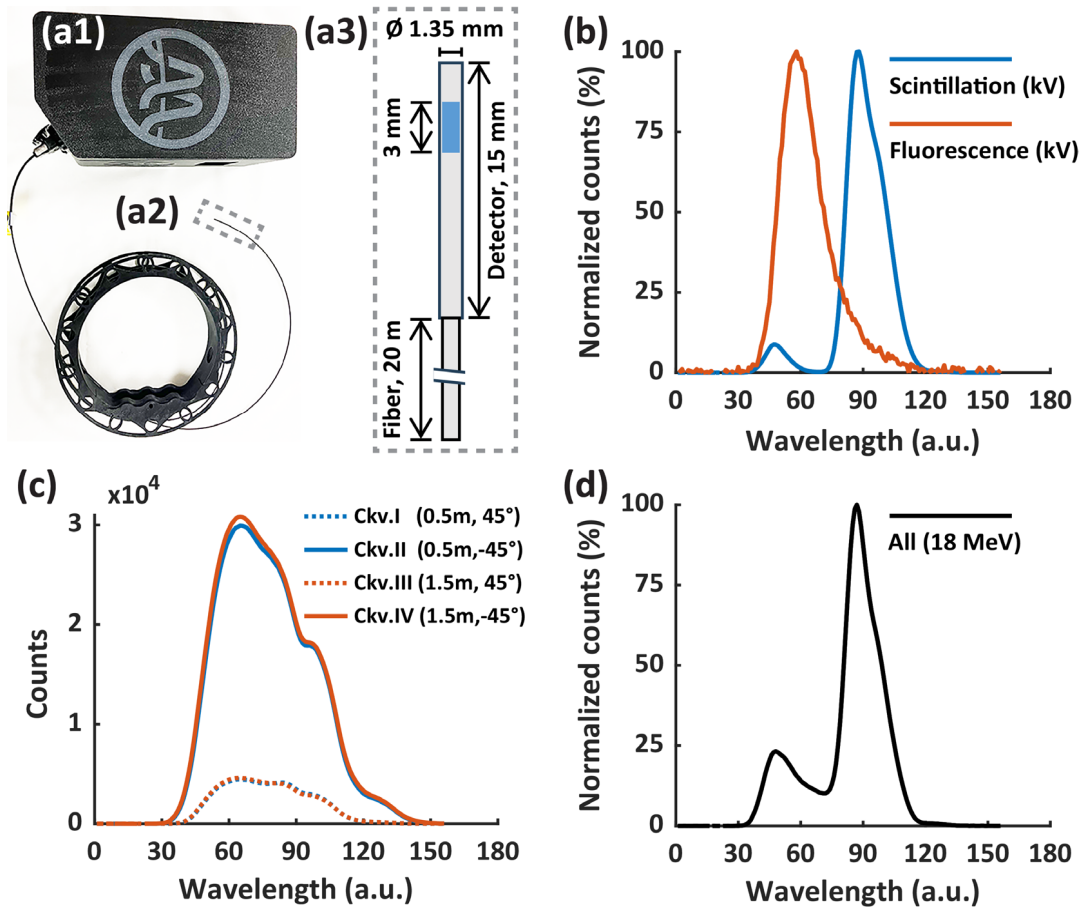

**Figure 1.** Configuration and calibration of the HYPERSCINT RP-FLASH scintillation dosimetry system. The HYPERSCINT RP-FLASH system comprises (a1) HYPERSCINT RP-FLASH optical reader, (a2) PRB-0042 RP detector, and HYPERDOSE software (not shown). (a3) illustrates the PRB-0042 RP dimension, sensitive volume (blue square, 1 mm diameter x 3 mm length) and optical fiber. (b) shows the spectrum of scintillation and fluorescence introduced by kV X-rays. (c) shows the spectrum of Cherenkov I&II (irradiated at 0.5 m from probe tip, ±45° gantry angle) and Cherenkov III&IV (irradiated at 1.5 m from probe tip, ±45° gantry angle) introduced by 6 MV X-rays. (d) is the spectrum after dose calibration at 18 MeV. Photodetector spectral response was not applied to the spectrum measurements. Abbreviations: a.u., arbitrary unit; Ckv.I-IV, Cherenkov I-IV.

The interaction of scintillator material with radiation results in the emission of optical photons, termed scintillation, and the number of scintillation photons is proportional to the absorbed dose. However, Cherenkov and fluorescence photons produced within both the scintillator and fiber optic would contaminate the scintillation signal, as they are not dose-proportional but related to other factors, such as the length of fiber optics irradiated. In the HYPERSCINT RP-FLASH system, the measured light signal is assumed to be a linear superposition of scintillation, fluorescence and Cherenkov photons. A hyperspectral approach was adopted to decompose the light components and calculate the dose with improved precision and robustness compared to other methods[35,36]. Dose and spectral calibrations are thus necessary for the hyperspectral approach.

Due to the absence of kV sources in the Clinac 21EX LINAC (Varian 21EX, Varian Medical Systems, Palo Alto, CA) used for dose calibration, the spectral calibration was conducted in a TrueBeam LINAC (Varian TrueBeam, Varian Medical Systems, Palo Alto, CA). Following the manual provided by the vendor (Medscint), a 4-component calibration was used for the spectral calibration, consisting of scintillation, fluorescence, Cherenkov I&II, and Cherenkov III&IV[35,37]. The scintillation component was acquired by irradiating the probe tip with minimal length of fiber optic by the kV imager of the TrueBeam, while the fluorescence component was characterized by irradiating 10 loops of the fiber optic while sparing the sensitive volume. The nominal energy of the kV beam is 90 kVp, lower than the Cherenkov light production energy threshold in the fiber optic. Figure 1b shows the scintillation and fluorescence spectrum measured by the HYPERSCINT system. Cherenkov I&II were acquired by irradiating the fiber optic at 0.5m from the probe tip by 6 MV of the TrueBeam, 100 cm source-to-surface distance (SSD), 40 x 40 $cm^2$ field and 6 cm solid water backscatter. Since the waveform of the Cherenkov signal is of primary interest, and is independent of beam modalities or energy[35,38], a 6 MV photon beam was chosen, as suggested by the vendor, using a large field setup to maximize the Cherenkov signal. Considering the maximum Cherenkov emission angle in PMMA of 47.8° [39], we used a 45° angle between the beam trajectory and the fiber axis to further enhance the Cherenkov signal. All the setups were the same for Cherenkov III&IV, except at 1.5m from the probe tip. Figure 1c shows the spectrum of Cherenkov I&II and Cherenkov III&IV. The rationale to irradiate different locations of the fiber with both 45° and -45° angle is to account for light attenuation in fiber optics and the relative direction between Cherenkov light transmission and fiber axis. Dose calibration was performed by delivering 500 monitor unit (MU) of 18 MeV CONV from the Clinac 21EX at 100 cm SSD, with a 10x10 $cm^2$ cone, at 2 cm depth, established by 1 cm solid water and 1 cm bolus, with 6 cm solid water as the backscatter piece (Fig. S1a). The probe tip was aligned to the crosshair of LINAC. This setup is referred to as the default setup. Since our animal FLASH experiment is conducted by 18 MeV UHDR, which provides superior dose homogeneity than lower energy electron beam, we therefore chose the 18 MeV beam as the default energy throughout this study. The 18 MeV CONV dose delivered to the scintillator probe was quantified by output measurement with an Accredited Dosimetry Calibration Laboratory (ADCL) calibrated

Semiflex 31013 IC (PTW-Freiburg, Freiburg, Germany) at 2 cm depth. Figure 1d is the normalized light spectrum resulting from the 18 MeV CONV irradiation.

The spectral calibration was recommended every 3 months by the vendor. The detector remained connected to the reader throughout the entire study to avoid potential calibration spectrum mismatch after unplugging and re-plugging the detector[33]. Dose calibration was performed prior to each set of measurements to ensure accuracy.

## 2.2 Film dosimetry

Gafchromic EBT-XD films (Ashland, Bridgewater, NJ) were used to validate the scintillator's UHDR dosimetric measurements, as the films have demonstrated dose rate independence at UHDR[7,40]. The films were scanned 24 hours post-irradiation using a flatbed photo scanner (Epson Expression 12000XL, Nagano, Japan) at consistent orientation. An in-house film analysis software was employed to extract dose information from the films, utilizing a dual-channel method (green and blue)[41]. The average readings in the central 5x5 $mm^2$ of films were used for the dosimetric analysis. The film dosimetry was also validated against IC measurements in CONV within 3% for the dose range of 0 to 40 Gy under the same setup.

## 2.3 CONV and UHDR measurements

Our group converted the Varian Clinac 21EX into an electron FLASH irradiator by configuring it in photon mode without the target and flattening filter, enabling maximum electron beam fluence and achieving 6 MeV UHDR irradiation[42]. Subsequently, the same approach was used to enable 18 MeV UHDR beam, achieving ~410 Gy/s at the isocenter with 2.3 Gy/pulse delivered at 180 Hz PRF. Detailed beam parameters are listed in Table S1. The corresponding relationship of MU, pulse number and dose output are shown in Fig. S2.

Unless otherwise stated, all CONV and UHDR measurements were performed using the default setup as aforementioned in Sec. 2.1. The range of study parameters and the default irradiation setting for CONV and UHDR beams used throughout this study are summarized in Table S1. For UHDR measurements, a piece of film was directly placed beneath the scintillator probe (Fig. S1b). Because the small-size probe was covered by the large 30 x 30 x 1 $cm^3$ bolus, the scintillator did not introduce dosimetric perturbation to film measurements.

Unless otherwise stated, all measurements reported in this study represent the average of three consecutive measurements, with error bars indicating the standard deviation of these measurements. The error bar for ratio analysis, such as the one in Fig. 2a2, represents its standard deviation calculated through propagation of uncertainty. Note that the error bars may not be visible if they are smaller than the symbols representing the average values.

## 2.4 Dose response and accuracy under CONV and UHDR irradiation

The dose linearity and dosimetric accuracy of the scintillator system were quantified for both CONV and UHDR beams at 18 MeV. For 18 MeV CONV, 1–3500 MU were delivered and the dose to the scintillator was quantified by the ADCL-calibrated Semiflex 31013 IC. Due to the long delivery time of CONV, 1 Hz $f_s$ was used. For 18 MeV UHDR, doses from 2–17 pulses were measured at 1 and 1000 Hz $f_s$ and verified against film measurements.

## 2.5 Beam energy, field size, dose per pulse, and pulse repetition frequency dependence

Given the water equivalence of PSDs, no energy dependence is expected for clinical electron and photon beams. However, due to continued interest from the scientific community and end users, we evaluated the energy dependence using commonly employed electron energies. The system's energy dependence was studied under CONV irradiation rather than UHDR and the reasons were twofold: first, IC measurements for CONV-RT provide less dosimetric uncertainties compared to film measurements for UHDR irradiation; second, study has shown the response of the scintillator system is independent of dose rate[34]. Therefore, 6 and 18 MeV CONV were used to assess the scintillator's energy response by evaluating whether the dose output measured by the scintillator differed significantly compared to the IC. The dose outputs of the 6 and 18 MeV CONV beams at the depth of scintillator position ($d_s$) quantified by IC ($D_{IC}$) were calculated by multiplying the daily outputs at $d_{max}$ measured by IC ($D_w^Q$), PDD, and water-to-solid water conversion factor ($C_{w \rightarrow sw}^Q$). The scintillator measured dose output ($D_{Scint}$) were then compared with the IC results. The standard deviation ($\sigma_{Rela}$, Eq. (1-1)) of the relative difference between scintillator and IC measurements ($Rela$) was calculated through uncertainty propagation. The $\sigma_{Scint}$ is the standard deviation of $D_{Scint}$. As shown in Eq. (1-2), the standard deviation of $D_{IC}$ ($\sigma_{IC}$) accounted for both the measurement uncertainty of IC ($\sigma_{Dw}$) and the uncertainty in PDD determination ($\sigma_{PDD}$) for output calculation, considering a depth placement uncertainty quantified as half the radius of the scintillator probe's sensitive volume (0.25 mm).

$$\frac{\sigma_{Rela}}{Rela} = \sqrt{\left(\frac{\sigma_{Scint}}{D_{Scint}}\right)^2 + \left(\frac{\sigma_{IC}}{D_{IC}}\right)^2} \qquad (1-1)$$

$$\frac{\sigma_{IC}}{D_{IC}} = \sqrt{\left(\frac{\sigma_{Dw}}{D_w^Q}\right)^2 + \left(\frac{\sigma_{PDD}}{PDD}\right)^2} \qquad (1-2)$$

Field size determines the length of fiber optics irradiated, thereby affecting the amount of fluorescence and Cherenkov light produced, a phenomenon known as the stem effect[15]. We assessed the field size dependence of the scintillator by comparing the field size factor ($f_{FS}$) measured by the scintillator to that measured by the IC. The $f_{FS}$ was defined as the ratio of dosimetric reading measured at a 2 cm depth with a specific field size to that measured with 10x10 cm$^2$ field. For both 6 and 18 MeV CONV beams, $f_{FS}$ for 6x6 to 25x25 cm$^2$ measured by the scintillator and IC (PTW Semiflex 31013) were compared. The chamber measurements were also verified with Varian golden beam data, showing within 1% difference (Fig. S3). Additionally, the $f_{FS}$ of the smaller field sizes 2x2 and 4x4 cm$^2$ were compared between scintillator and CC13 ion chamber (3 mm radius x 5.8 mm length cavity, IBA dosimetry, Schwarzenbruck, Germany).

To investigate the DPP dependence of the scintillator system for UHDR electron irradiation, we adjusted SSD from 100 to 150 cm at 5 cm increment for our scintillator setup, which in turn altered the dose output and thus DPP measured at a given SSD. Ten 18 MeV UHDR pulses were delivered at 100–150 cm SSD with 180 Hz PRF (Table S1). EBT-XD film was used to assess the relationship between the dose output at UHDR versus SSD and if it follows the inverse square law. DPP at various SSDs can then be determined by the film-measured output-SSD relationship and the number of pulses.

We further examined if the system's response depends on the PRF from 30 to 180 Hz by delivering 4 to 17 pulses to the scintillator, with a DPP of 2.3 Gy. The PRF was adjusted by setting the repetition rate from 100 to 600 MU/min in the LINAC console (Fig. S4). The PRFs of the delivered beams were verified by a remote trigger unit (RTU; DoseOptics, Lebanon, NH) positioned outside the radiation field (Fig. S4). The RTU is a coincidence-based radiation detector; it consists of two scintillators and corresponding silicon photomultipliers (SiPMs) to detect scattered radiation[43].

## 2.6 Radiation degradation

To quantify the radiation degradation of the scintillator and fiber optic, the detector degradation factor (DDF) was evaluated as a function of the accumulated dose:

$$DDF(D_a) = \frac{Rdg(D_a)}{Rdg(D_0)} \qquad (2)$$

where $D_a$ is the accumulated dose, defined as the dose delivered to the scintillator without dose recalibration performed. $Rdg(D_a)$ represents the scintillator's reading after measuring 100 MU of 18 MeV CONV using the default setup at an accumulated dose of $D_a$. On the other hand, $Rdg(D_0)$ is the scintillator's reading under the same condition acquired immediately after calibration, when the radiation-induced degradation affecting the scintillator reading is eliminated.

A total $D_a$ of around 10 kGy was delivered to the scintillator within 2 hours using 18 MeV UHDR. Since the scintillator's reading may be subject to uncertainty due to the radiation degradation, an independent detector is required to determine $D_a$ delivered to the scintillator. A CC13 IC was placed under the edge of the electron cone (Fig. S5a) to measure the Bremsstrahlung and scattered radiation. Due to the relatively small sensitive volume of CC13 (0.13 cm$^3$) and the low dose rate of Bremsstrahlung and scattered radiation, the CC13 reading has been shown to be linear with the FLASH dose delivered to the scintillator (Fig. S5b). The accumulated dose $D_a$ to the scintillator can thus be quantified based on the CC13 reading.

## 2.7 Verification of pulse-resolved dose measurements at UHDR

To verify the dosimetric accuracy of the scintillator system in measuring inter-pulse delivery, the scintillator-measured dose for a given pulse or DPP along a train of 10 pulses delivery at 180 Hz PRF was compared against the measurements from a pulse form monitoring device. The device consists of an HC-120 series photomultiplier tube (PMT; Hamamatsu, Shizuoka, Japan) coupled with an optical fiber of 1 mm diameter, positioned 3 m from the isocenter to detect scattered radiation through radiation-induced Cherenkov emission[24]. The PMT output was fed into an oscilloscope (70MHz, PicoScope 3000 series, Pico Technology, Cambridgeshire, UK), allowing us to monitor the LINAC pulse amplitudes on a pulse-by-pulse basis. The normalized area under curve (AUC) of the pulse form was then compared to the normalized DPP measured by the scintillator. The scintillator was operated under 1000 Hz *fs*. Ten repeated measurements were performed for both the scintillator and PMT.

## 3. RESULTS

### 3.1 Dose response and accuracy under CONV and UHDR irradiation

The performance of the scintillator in response to 18 MeV CONV and UHDR was benchmarked against IC and film measurements, respectively. As shown in Fig. 2a1, the scintillator at 1 Hz *fs* demonstrated good linearity compared to IC measurements for 18 MeV CONV up to 35 Gy. The scintillator achieved an accuracy within 0.5% compared to the IC measurements from 0.1 to 35 Gy, and up to 4% deviation was seen for dose <0.1 Gy (Fig. 2a2).

For 18 MeV UHDR, Fig. 2b1 shows that the scintillator at 1 Hz *fs* displayed reasonable linearity up to ~18 Gy, beyond which saturation was observed. This behavior was consistent with the spectrum measured after 4 pulses (9 Gy) and 17 pulses (40 Gy) delivered (Fig. 2b2).

It renders that the nonlinear response observed at 1 Hz *fs* likely resulted from large amount of light signal collected within a sampling window exceeding the dynamic range of the photodetector array during UHDR irradiation. By increasing the sampling frequency to 1000 Hz, we were able to resolve the saturation issue. Compared to the film measurements, the scintillator at 1000 Hz *fs* showed good dose linearity ($R^2 > 0.999$) and within 3% accuracy up to 40 Gy (Fig. 2c1-2).

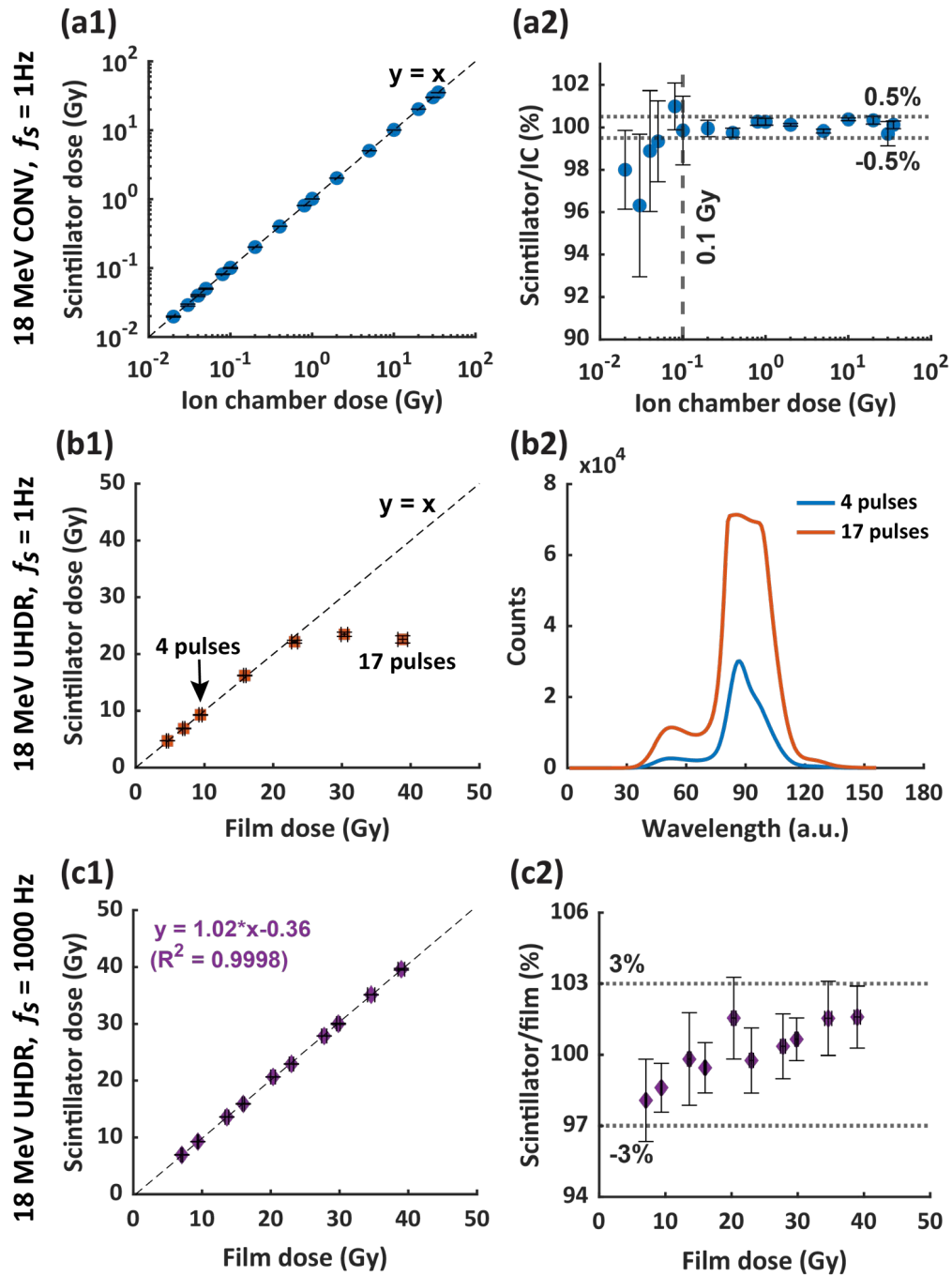

**Figure 2.** Dose linearity and accuracy of the scintillator under CONV and UHDR electron irradiation at different sampling frequencies. (a1-2) show the readings of the scintillator at 1 Hz *fs* compared to IC measurements and the ratio of the measurements under 18 MeV CONV. (b1) shows the readings of the scintillator at 1 Hz $fs$ against EBT-XD film under 18 MeV UHDR; (b2) is the corresponding light spectrum measured by the scintillator after receiving 4 and 17 pulses of 18 MeV UHDR. (c1-2) show the dose measured by scintillator at 1000 Hz $fs$ compared to that of film and their ratio under 18 MeV UHDR. Abbreviations: $f_s$, sampling frequency; CONV, conventional dose rate; UHDR, ultra-high dose rate; IC, ionization chamber; a.u., arbitrary unit.

## 3.2 Beam energy, field size, dose per pulse, and pulse repetition frequency dependence

Table 1 compares the scintillator and IC measurements after receiving 100 MU of 6 and 18 MeV CONV. The relative differences between the scintillator and IC measurements were 1.81±2.18% for 6 MeV and 0.18±0.13% for 18 MeV. The 0.25 mm scintillator placement uncertainty (Sec. 2.5) primarily contributed to the large variation in the PDD determination for 6 MeV, resulting in a larger $Rela$ seen in the 6 MeV case compared to 18 MeV. From two-tailed t-test, the detectors' relative reading differences between 6 and 18 MeV have $p = 0.325$, indicating no significant energy dependence.

**Table 1.** Energy dependence of scintillator for 6 and 18 MeV CONV compared to IC measurement

| Energy | $d_s$ (cm) | $D_{Scint}@d_s$ (cGy) | Ion chamber | | | | $Rela$ |
|---|---|---|---|---|---|---|---|
| | | | $D_{w,dmax}^{Q}$(cGy) | $PDD(d_s)$ | $C_{w \to sw,ds}^{Q}$ | $D_{IC}@d_s$(cGy) | |
| 6 MeV | 2±0.025 | 85.36±0.14 | 100.79±0.10 | 81.75±1.74% | 1.0176 | 83.84±1.79 | 1.81±2.18% |
| 18MeV | 2±0.025 | 100.81±0.11 | 101.55±0.05 | 99.80±0.03% | 0.9929 | 100.63±0.06 | 0.18±0.13% |

Abbreviations: $D_{Scint}$, output measured by scintillator; $D_{IC}$, output measured by ion chamber; $d_s$, depth of scintillator position; $D_{w,dmax}^{Q}$, the absorbed dose to water at beam quality $Q$ at $d_{max}$ and 100 MU delivered; $PDD(d_s)$, percent depth dose at $d_s$; $C_{w \to sw,ds}^{Q}$, water to solid water conversion factor for beam quality $Q$ at depth $d_s$; $Rela$, the relative difference between the scintillator and IC measurements.

Figure 3a shows the field size factors $f_{FS}$ of 6 and 18 MeV CONV measured by the scintillator and IC from 2x2 to 25x25 cm². In contrast to 18 MeV, $f_{FS}$ of 6 MeV rapidly increased from 2x2 to 4x4 cm² and reached saturation, partly due to the sharp drop of the depth dose at the 2 cm depth of measurement for the 2x2 cm² field. Figure 3b illustrates the ratio of $f_{FS}$ measured by the scintillator to those measured by the IC. The scintillator and IC measurements agreed within 1.5% accuracy, indicating that the scintillator exhibits <1.5% field size dependence from 2x2 to 25x25 cm² for both energies.

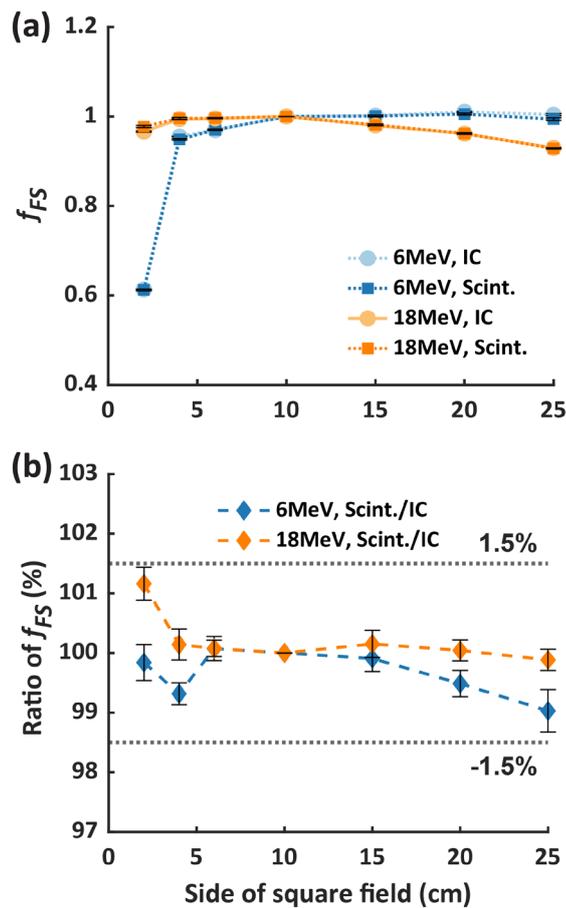

**Figure 3.** Filed size dependence of the scintillator under 6 and 18 MeV UHDR irradiation. (a) shows the field size factor $f_{FS}$ measured by the scintillator and IC, and (b) is the corresponding $f_{FS}$ ratio. Abbreviations: $f_{FS}$, field size factor; IC, ion chamber; Scint., scintillator.

Figure 4a shows that the square root of the film dose ratio between 100 cm and other SSDs exhibited good linearity ($R^2 > 0.999$), indicating an inverse square relationship between dose output and SSD for our 18 MeV UHDR beam. The DPPs ranging from 0.8 to 2.3 Gy/pulse at 100–150 cm SSD were then calculated based on the output-SSD relationship (Fig. 4a) and the number of pulses. Figure 4b shows the relative scintillator reading versus DPP, normalized at 2.3 Gy/pulse, demonstrating good linearity ($R^2 > 0.998$). The ratio of scintillator-to-film doses versus DPP shows <2% variation across 0.8–2.3 Gy/pulse (Fig. 4c), indicating minimal DPP dependence.

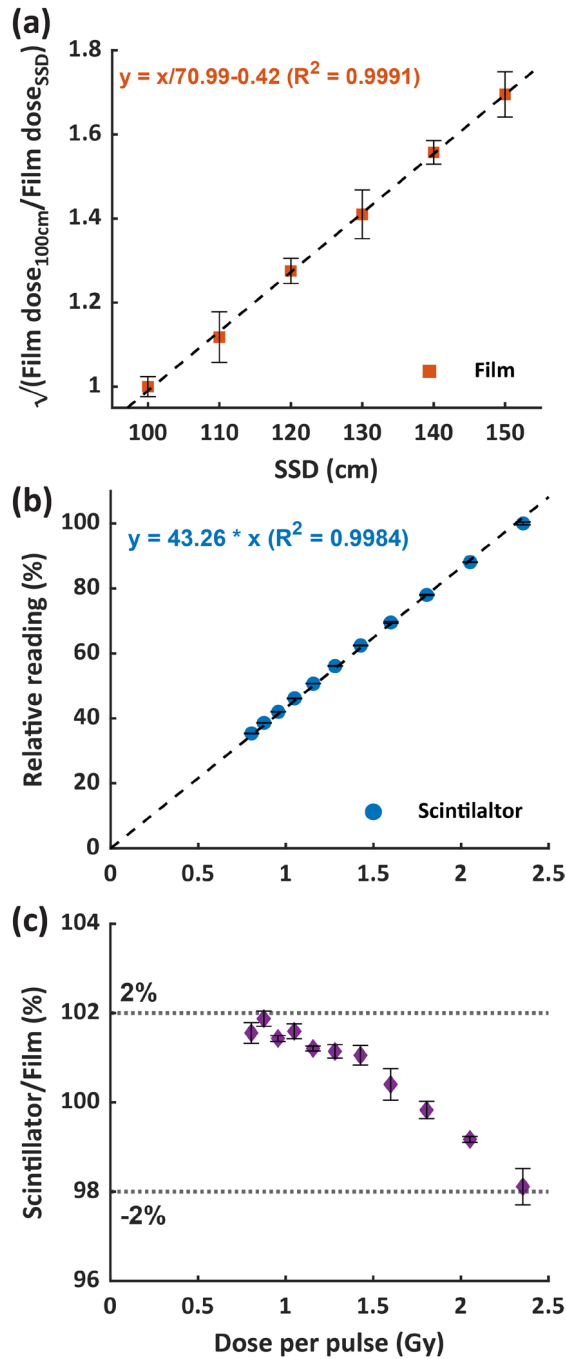

**Figure 4.** Dose per pulse dependence of the scintillator under 18 MeV UHDR irradiation. (a) shows the square root of the ratio of dose outputs measured by film between 100 cm and other SSDs as a function of SSD. (b) shows the normalized scintillator reading after ten 18 MeV UHDR pulses delivered with respect to DPP, normalized at 2.3 Gy/pulse. (c) is the ratio of the scintillator-to-film doses vs DPP. Abbreviation: DPP, dose per pulse; Film dose$_{100cm}$\Film dose$_{SSD}$, the ratio of dose output measured by film at 100 cm SSD vs other SSDs.

We analyzed the impact of PRF on scintillator measurements by delivering 4 to 17 pulses of 18 MeV at UHDR, with PRF ranging from 30 to 180 Hz (Fig. 5). For the 4 pulses scenario with 9 Gy delivered, minimum PRF dependence (within 1%) was observed. For 10 to 17 pulses delivering 23–40 Gy, the normalized readings decreased as the PRF increased, though the variation among PRF settings was less than 2%.

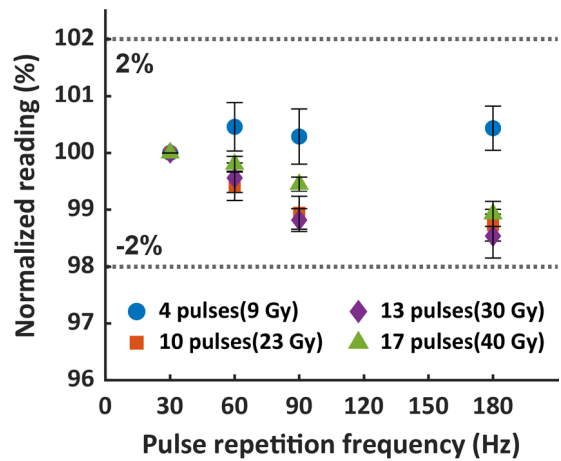

**Figure 5.** Normalized scintillator reading versus 30–180 Hz PRF after receiving 4–17 pulses of 18 MeV at UHDR (9–40 Gy). Abbreviation: PRF, pulse repetition frequency.

### 3.3 Radiation degradation

Figure 6a shows the detector degradation factor (DDF, see Sec. 2.6) as a function of the accumulated dose. The DDF can be modeled by a 2nd-order polynomial fit up to 10 kGy. We provide a quick estimation of -2.65%/kGy degradation after 2 kGy delivered to scintillator, which can serve as a guide for dose recalibration. Figure 6b illustrates the measured light spectrum with respect to the accumulated dose. Detector degradation was observed to result not only in a reduction of the scintillator signal but also in changes to the light spectrum, as evidenced by the diminishing small peak with increasing accumulated dose.

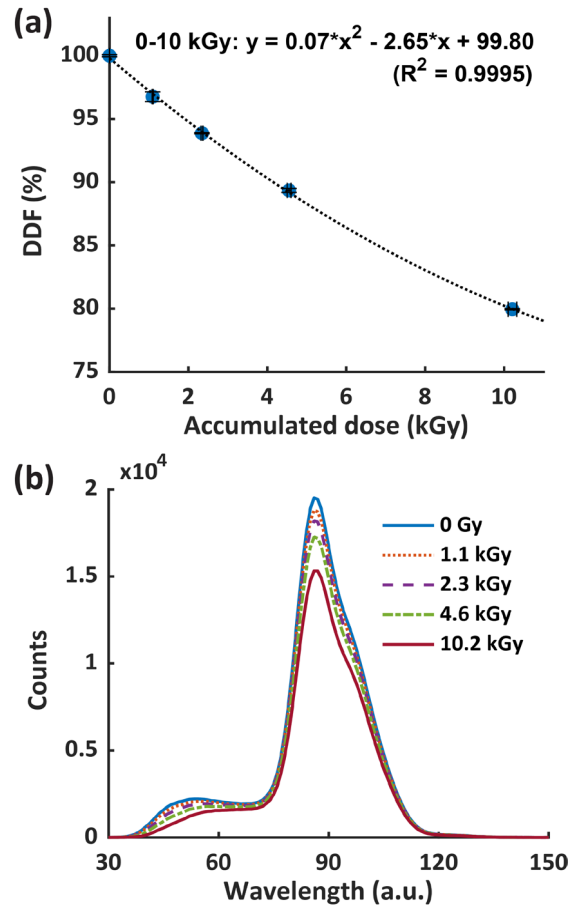

**Figure 6.** (a) is DDF of the scintillation detector as a function of accumulated dose, and (b) is the associated total light spectrum, including scintillation, fluorescence and Cherenkov, with respect to the accumulated dose. Abbreviation: DDF, detector degradation factor; a.u., arbitrary unit.

### 3.4 Verification of pulse-resolved dose measurements at UHDR

Figure 7a shows a representative PMT-fiber optic measurement detecting scattered radiation from a given radiation pulse. The time interval between the PMT signal onset and its peak corresponds to the radiation pulse width, while the subsequent signal decay reflects the PMT system response. The AUCs of both the full PMT waveform and the segment from onset to peak exhibit a linear relationship with the delivered dose measured by film ($R^2 > 0.999$, Fig. S6), indicating that scattered radiation is proportional to the delivered dose. Figure 7b shows the normalized pulse-resolved readings of the scintillator and PMT for a train of 10 pulses 18 MeV UHDR delivery. The reading of the scintillator refers to the measured DPP, while the reading of the PMT represented the AUC of the whole detected signal in response to each pulse (Fig. 7a). The pulse-resolved readings from both detectors were normalized to each detector's average reading of the 10 pulses, respectively. The scintillator and PMT readings both showed a consistent lower amplitude for the 1st pulse and a higher amplitude for the 2nd one, while for the remaining pulses, the results were relatively stable. Figure 7c further illustrates that the ratio between the scintillator and PMT readings agrees within 3%. A large standard deviation of the ratio is attributed to PMT signal uncertainty (Fig. 7b), which may arise from its high sensitivity making the signal susceptible to gain voltage, secondary electron emission, and environmental factors such as thermionic emission and temperature. Additional signal fluctuations can result from weak scattered radiation detected by the PMT positioned 3 m from the isocenter and the small sensitive volume of the 1 mm diameter fiber optic. Nevertheless, our results confirm the applicability of the scintillator for pulse-resolved dose measurements at UHDR electron irradiation.

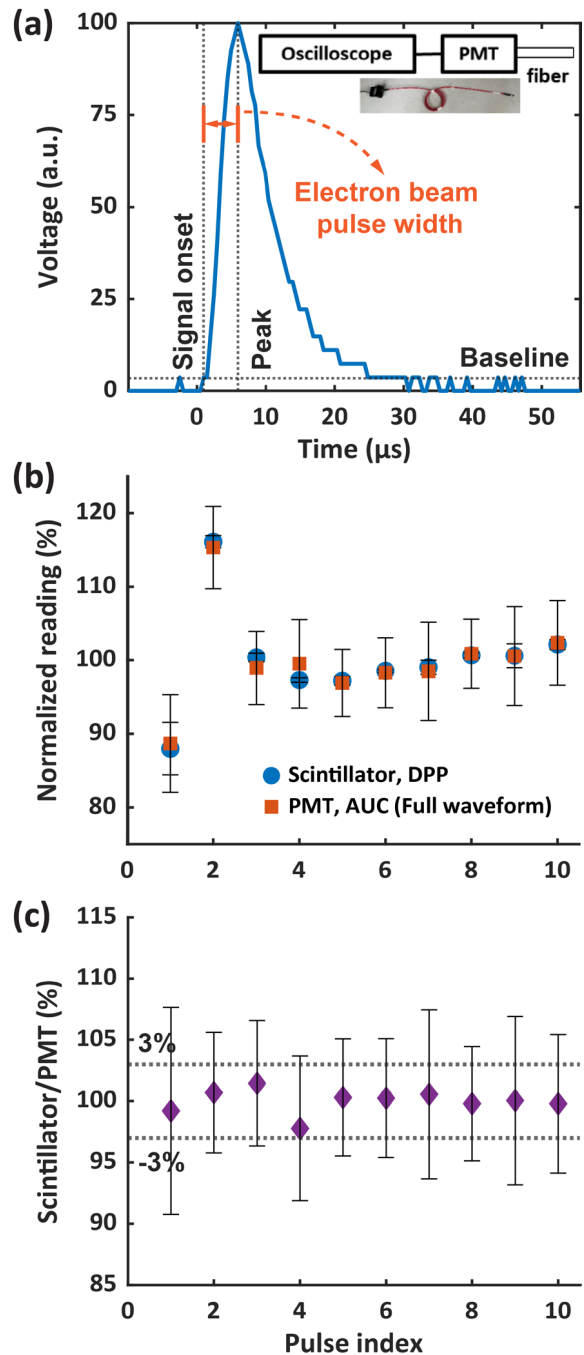

**Figure 7.** Comparison of the scintillator and PMT-fiber optics for pulse-resolved dose measurements. (a) shows the configuration of the scattered radiation-based pulse form monitoring unit, which consists of a PMT coupled with an optical fiber, and its measured waveform after receiving one 18 MeV UHDR pulse. (b) shows the comparison of the relative pulse-resolved reading of the scintillator (DPP) and PMT (AUC of the full waveform, Fig. a) for ten 18 MeV UHDR pulses, normalized to the average reading of the 10 pulses. (c) shows the corresponding ratio of the scintillator and PMT measurements. Abbreviations: a.u., arbitrary unit; AUC, area under curve; DPP, dose per pulse; PMT, photomultiplier tube.

## 4. DISCUSSION

The introduction of UHDR radiation presents substantial dosimetric challenges, particularly in accurately measuring radiation dose under conditions of high dose per pulse within an extremely short period of time. Scintillators have emerged as detectors of choice to address these challenges[30-34,44-46]. Our study provides a comprehensive characterization of the HYPERSCINT RP-FLASH scintillator, evaluating its suitability as a dosimeter for UHDR electron research.

The HYPERSCINT RP-FLASH scintillator system exhibited high accuracy in dose measurements for both CONV (~0.1 Gy/s) and UHDR (~410 Gy/s) electron irradiation (Fig. 2). With an accuracy within 0.5% over the dose range of 0.1 to 35 Gy in CONV mode, the system aligns with the performance of previous HYPERSCINT versions at extended dose ranges[19,47]. However, nonlinear behavior in scintillator response was observed at the 1 Hz $f_s$ when doses <0.1 Gy or >18 Gy were delivered. These nonlinearities are likely due to the photodetector's signal processing limitations: at low doses, measurement accuracy is compromised by system noise perturbing the weak light signal, while at high doses, the light signal exceeds the detector's dynamic range (Fig. 2b2). Reducing the system's gain may be a potential way to mitigate the saturation issue. However, this approach requires re-calibration, as altering the gain affects both dose and spectral calibration. Additionally, it reduces the signal amplitude and decreases the signal-to-noise ratio, which can negatively impact measurement accuracy. Increasing the sampling frequency, which shortens the sampling window and reduces the dose measured per window, effectively mitigates the saturation effect observed at higher doses. At 1000 Hz $f_s$, the scintillator achieved <3% dose variation compared to the film measurements up to 40 Gy, covering the dose range used in most FLASH studies. Since direct reference dosimetry for UHDR electron beams traceable to metrological standards is currently unavailable, EBT-XD film, with a dosimetric accuracy ~3%, was used as a substitute. The precision of film in dose measurement thus limits the upper bound of our assessment of scintillator performance. We expect the actual accuracy of the RP-FLASH scintillator in UHDR dosimetric measurement to be better than 3%.

Evaluating dose rate linearity is important for UHDR dosimeters. Baikalov et al.[34] demonstrated that the RP-FLASH system exhibited dose rate independence from 1.8 to 1341 Gy/s. Since their study already spanned a wide range of dose rates range beyond the operating limits of our electron FLASH LINAC system, we did not include the dose rate linearity characterization in this study.

The scintillation signal from the scintillator is independent of the incident particle's energy above ~125 keV[15]. Additionally, the spectra of Cherenkov and fluorescence emissions produced in PMMA fiber are indistinguishable across energies from $^{60}$Co to 15 MeV[38]. Given the insensitivity of the light spectrum to beam energy and the water equivalence of scintillators, significant beam energy dependence is not expected for PSDs used with clinical electron and photon beams. The RP-FLASH scintillator demonstrated <2% dependence between 6 and 18 MeV (Table 1), further confirming no significant energy dependence across the commonly used electron beams.

The stem effect poses a challenge in scintillator dosimetry, as it can introduce errors in dose measurements[15]. Field size can influence the stem effect by determining the volume of fiber irradiated and, consequently, the intensity of Cherenkov and fluorescence light. Beierholm et al.[48] reported that the relative difference of Exradin W1 to IC measurements are up to 3.3% for 2x2–10x10 cm$^2$ fields and remain within ~0.8% for 10x10–25x25 cm$^2$ at 6–15 MV of Varian TureBeam LINAC. Recently, the Exradin W2 system has been well-characterized in small field dosimetry. Al Kafi et al.[18] and Okamura et al.[21] showed that the W2 system exhibited <2 % field size dependence for 0.5–5 cm diameter circular field using 6 MV CyberKnife beam. Jermain et al.[20] confirmed the <2% field size dependence for 0.4–2.5 cm diameter circular filed for ZAP-X 2.7 MV beam. However, to the best of our knowledge, the field size dependence of W2 for larger field sizes has not been reported. In our study, we evaluated the field size dependence of the RP-FLASH scintillator across both small and large fields, demonstrating less than <1.5% deviation across 2x2 to 25x25 cm$^2$ for 6 and 18 MeV (Fig. 3b). These results suggest that the hyperspectral method effectively mitigates the stem effect by addressing variations in Cherenkov and fluorescence light intensity as a function of field size.

Dose per pulse is a known limiting factor for the performance of scintillators. Since tuning pulse width is not feasible with our electron FLASH platform, we modified DPP by varying SSD and concurrently instantaneous dose rate to assess its effect on scintillator performance.

This method has been used in previous studies[31,34] to evaluate the DPP dependence of PSDs. Liu et al.[31] observed that the Exradin W2 scintillator exhibits saturation at DPP levels exceeding 1.5 Gy. The W2 system paired with the MAX SD unit effectively resolves this issue, demonstrating less than 2% DPP dependence up to 3.6 Gy[32]. Baikalov et al.[34] reported within 2% signal decrease for the same RP-FLASH scintillator at 0.4–2.2 Gy DPP and a 2–6% decrease for 2.2–7.7 Gy. Our findings of DPP dependence within 2% over the range of 0.8–2.3 Gy (Fig. 4) are consistent with the study of Baikalov et al.[34]. Adjusting SSD with the same electron cone inherently alters the field size. In this study, the 10x10 cm$^2$ cone at 100–150 cm SSDs corresponds to projected field sizes of 10x10–15x15 cm$^2$ defined at surface. From Fig. 3b, the scintillator only exhibited <0.2% variation between 10x10–15x15 cm$^2$, suggesting that field size would not play significant factor to the observed DPP dependence of the RP-FLASH scintillator (Fig. 4b).

Pulse repetition frequency may influence scintillator performance in UHDR dosimetry, particularly at high DPP levels. Liu et al.[31] investigated the impact of PRF 10–120 Hz on the signal response of Exradin W2 scintillator for DPPs from 1.2 to 7.3 Gy/pulse using 9 MeV UHDR electrons. They observed that the W2 signal response across PRFs of 10–90 Hz for the blue channel and 10–120 Hz for the green channel remained within 1.5% of the signal at 10 Hz for DPPs between 1.2 to 2.7 Gy/pulse. However, when the DPP increased to 7.3 Gy/pulse, at 120 Hz PRF, the W2 response dropped by as much as 30 and 20% for the blue and green channels, respectively. The authors hypothesized that this dependence was due to limitations in the electrometer's signal processing capabilities. After paired with the MAX SD optical detection and signal processing unit, Oh et al.[32] reported the W2 system exhibited no significant change in the collected charge across 18–180 Hz PRF after 10 pulses of 16 MeV UHDR delivery, although DPP information was not provided. The HYPERSCINT system uses a photodetector array as the readout unit, exhibiting <2% dependence on PRFs between 30 and 180 Hz after 4–7 pulses of 18 MeV UHDR beam delivered at 2.3 Gy/pulse (Fig. 5). Although the DPP we evaluated is a subset of that used by Liu et al.[31], our current findings support the signal response of HYPERSCINT system is not suspectable to PRF variation ranging from 30–180 Hz.

Plastic scintillators have demonstrated both permanent and temporal signal degradation due to radiation exposures[49]. The signal recovery due to the temporal degradation can occur over ~2 months[50]. Notably, higher dose rate irradiation introduces less permanent damage to the scintillation detector thus more temporal degradation[49]. Recent studies have examined the radiation damage to plastic scintillators used in UHDR dosimetry[30,31,34], with findings ranging from no evident signal degradation to 16%/kGy. However, the dose rates in these studies were not reported explicitly, complicating the interpretation of results. Recently, Giguère et al.[50] delivered 37.2 kGy of 200 MeV electron beam at fixed ~40 Gy/s to a polystyrene based scintillator (Medscint Inc. Quebec, Canada) and reported a 1.55%/kGy light output loss. In our study, ~10 kGy of 18 MeV electrons were delivered at ~410 Gy/s over a period of 2 hours, considerably shorter than the reported 2 months recovery period[50]. We observed a decrease in the DDF that followed a 2nd-order polynomial trend, with a linearly fitted degradation rate of 2.65%/kGy for 0–2 kGy (Fig. 6a). The difference between the degradation rates of 1.55%/kGy and 2.65%/kGy likely arises from the dose rate difference (~40 Gy/s vs. ~410 Gy/s), which may lead to different levels of temporal damage. A smaller degradation rate and DDF are expected for the RP-FLASH scintillator after the detector resolved from the temporal damage. Our results also indicate that the rate of signal degradation decreases with accumulated radiation dose (Fig. 6a). Before the degradation assessment, our scintillator detector had been exposed to radiation. It is reasonable to infer that an unirradiated scintillator would exhibit a higher signal degradation rate. Additionally, we demonstrated that radiation degradation of the scintillation detector is associated with both signal reduction and changes in the scintillator's light spectrum (Fig. 6b), consistent with the findings of Giguère et al.[50]. Based on our results, we recommend a conservative practice; performing dose recalibrations prior to measurements and whenever the accumulated dose reaches 1 kGy to ensure dose measurement accuracy within 3%.

Radiation degradation in scintillation detectors arises from damage to both the scintillator and fiber optic components. Adopting red-shifting dopants to plastic scintillators may enhance the radiation hardness, as radiation-induced light transmission losses occur primarily around 400–600 nm, while remaining relatively stable in the 600–800 nm range[30]. Additionally, employing high $OH^-$ content pure silica fiber optics, which exhibit superior radiation resistance[51], can mitigate radiation-induced fading of fiber optics.

In addition to dose and average dose rate, dose per pulse has been shown to influence the FLASH effect[13], emphasizing the need for a pulse-resolved UHDR dosimeter. Beam current transformer[52] and PMT-fiber optic detector[24] can measure time-resolved dosimetric parameters in a relative manner and are capable of pulse-resolved measurement. Previous studies have demonstrated that scintillators can provide pulse-resolved dose measurements[30,33], however no other time-resolved dosimeters were used simultaneously in those studies to verify the measurement results. In this study, we verified the accuracy of RP-FLASH scintillator for resolving dose from single pulse against the PMT-fiber optic measurements within 3% deviation. Given its 1kHz sampling frequency, the scintillator system can quantify inter-pulse dosimetry including pulse number, pulse-resolved DDP, and average dose rate of UHDR electron beam. However, the RP-FLASH scintillator is unable to measure pulse width or intra-pulse dose rate because the 4.5 μs LINAC pulse duration is considerably shorter than the 1 ms sampling window of the scintillator system. This limitation stems from the 1000 Hz sampling frequency of the RP-FLASH optical reader, not from the scintillator itself. Replacing the optical reader with an alternative system capable of sampling frequencies above 10 MHz, such as a photodiode integrated with customized electronics[53], could enable intra-pulse measurements. Solid state detectors have also been investigated in intra-pulse measurements. When paired with a 200 MHz oscilloscope, the flashDiamond detector (type 60025, PTW-Freiburg, Germany) can achieve sub-μs resolution, allowing measurement of pulse width and instantaneous dose rate of individual beams[54].

## 5. CONCLUSIONS

This study comprehensively investigated the performance of the HYPERSCINT RP-FLASH scintillator system for electron UHDR research. The scintillator system demonstrated good dose linearity and accuracy for both CONV and UHDR electron beams, with minimal dependence on beam energy (6 and 18 MeV), field size (2x2–25x25 cm$^2$), dose per pulse (0.8–2.3 Gy), and pulse repetition frequency (30–180 Hz). The radiation degradation of the scintillation detector followed a 2$^{nd}$-order polynomial fitting, with -2.6%/kGy for 0–2 kGy. The system was also capable of performing pulse-resolved UHDR dose measurements at 1000 Hz sampling frequency. Our study suggests that the HYPERSCINT RP-FLASH scintillator system could be a suitable detector for UHDR electron dosimetry. This work thus provides the research community with guidance on utilizing this commercially available scintillator system for electron FLASH research.

## ACKNOWLEDGEMENTS

The authors acknowledge the funding support from Cancer Prevention and Research Institute of Texas, RR200042. They also thank François Therriault-Proulx, Benjamin Côté and Danahé LeBlanc (MedScint, Quebec City, Canada) for technical support and useful discussion.

## CONFLICTS OF INTEREST STATEMENT

The authors have no relevant conflicts of interest to disclose.

## DATA AVAILABILITY STATEMENT

Data are available from the corresponding authors upon reasonable request.